\documentclass[aps,amsmath,amssymb,amsfonts,twocolumn]{revtex4-2}
\usepackage{graphicx}
\usepackage[colorlinks=true,linkcolor=blue,citecolor=blue, urlcolor=blue]{hyperref}
\usepackage{blindtext}

\newcommand{\RNum}[1]{\uppercase\expandafter{\romannumeral #1\relax}}
\newcommand{\RNumLow}[1]{\lowercase\expandafter{\romannumeral #1\relax}}

\usepackage{lineno}

\usepackage{setspace}
\AtBeginDocument
{
	\addtolength\abovedisplayskip{0.20\baselineskip}%
	\addtolength\belowdisplayskip{0.20\baselineskip}%
}

\begin{document}

\title{pH modulates friction memory effects in protein folding}

\author{Benjamin A. Dalton}
\affiliation{Freie Universit\"at Berlin, Fachbereich Physik, 14195 Berlin, Germany}
\author{Roland R. Netz\thanks{Corresponding author}}
\affiliation{Freie Universit\"at Berlin, Fachbereich Physik, 14195 Berlin, Germany}

\begin{abstract}
We study the non-Markovian folding dynamics of the $\alpha$3D protein under low- and neutral-pH conditions. Recently published all-atom simulations of $\alpha$3D by the Shaw group reveal that lowering the pH significantly reduces both native and non-native salt-bridge interactions, which dominate the folding dynamics. Here, we demonstrate that this physiochemical modulation directly perturbs the folding friction, which we evaluate using non-Markovian memory-kernel-extraction techniques. In doing so, we find that the reduction in pH not only decreases the magnitude of the time-dependent friction acting on the protein but also more dramatically shortens the time scale of the friction memory effects. As a result, the folding dynamics in the low pH system are well described by a purely Markovian model. In the neutral pH system, however, the memory time scale is of the same order as the folding time and is accelerated by a factor of 6 compared to a Markovian model prediction. We demonstrate that this memory-induced barrier-crossing speed-up is predicted by non-Markovian reaction-kinetic theories, confirming that non-Markovian models are, in general, necessary for a quantitative description of protein folding dynamics.
\end{abstract}

\maketitle

Protein folding is an essential requirement for many biological functions and has long been the interest of theoretical and computation studies \cite{Camacho_1993, Bryngelson_1995, Dill_1997, Plotkin_2003, Sosnick_2011,Hinczewski_2013}. To describe the folding dynamics of a protein, one typically represents an instantaneous protein structure with a suitably chosen one-dimensional reaction coordinate. Hence, the time evolution of the reaction coordinate is modelled as a diffusive process over the extracted free energy profile. This coarse-graining naturally incurs friction, representing the dissipative interactions between the reaction coordinate and its environment. Evaluating the friction acting on an arbitrary reaction coordinate is a non-trivial task and depends on the model used to describe the reaction coordinate dynamics. Markovian models have been successfully applied to model protein folding \cite{Munzo_1999, Best_2006, Zheng_2015b}. However, these models neglect non-Markovian effects, suggesting that such effects are fast compared to the relevant folding time scales. Two problems with this approach are that \RNumLow{1}) it has been shown that non-Markovian effects can still influence barrier-crossing kinetics when memory time scales are short compared to barrier-crossing time scales \cite{Kappler_2018}, and \RNumLow{2}) proteins do exhibit long memory time scales \cite{Dalton_2023}. While non-Markovian effects have been considered in the context of protein folding \cite{Plotkin_1998, Berezhkovskii_2018}, recent memory kernel extraction techniques \cite{Berne_1970, Daldrop_2018, Kowalik_2019, Ayaz_2021} have enabled the accurate evaluation of time-dependent friction kernels acting on arbitrary reaction coordinates and have been validated for a range of molecular systems \cite{Brunig_2022a, Brunig_2022d, Dalton_arXiv_2023}, thus enabling a direct method to evaluate the relevance of non-Markovian effects.

Recently, we applied non-Markovian memory kernel extraction techniques to study the folding and unfolding dynamics of a set of all-atom, fast-folding protein simulations. We showed that all proteins in the study exhibit significant non-Markovian dynamics when described using a well-characterized fraction of native contacts reaction coordinate \cite{Dalton_2023}. We showed that memory decay times are comparable to the folding time scales and that, overall, non-Markovian processes are not fast compared to the relevant protein folding kinetics. Furthermore, we also showed that reaction rate theories that explicitly incorporate non-Markovian effects are successful at predicting folding and unfolding times across the full population of proteins. In the present article, we apply non-Markovian techniques to study of the folding of a small, fast-folding protein under neutral and perturbed pH conditions, and we show that the observed deviations away from a Markovian prediction for the folding kinetics of the neutral pH protein are as expected according to various non-Markovian predictions for the folding kinetics of proteins.

\begin{figure}[b!]
\includegraphics[scale=1.05]{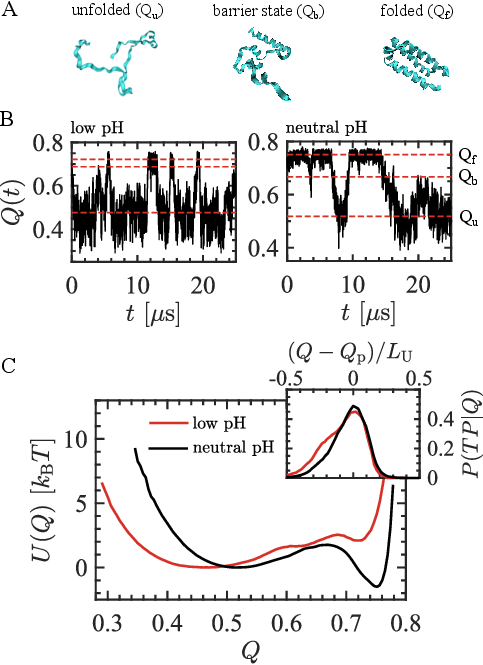}
\caption{
Folding dynamics of the $\alpha$3D protein. A) Simulation snapshots of representative states for neutral pH. All-atom trajectories taken from \cite{Lindorff_2011, Chung_2015}. B) 25 $\mu$s trajectory segments for the fraction of native contacts $Q$ reaction coordinate for $\alpha$3D for low pH (pH=2) and neutral pH (pH=7). The unfolded, barrier top, and folded states are indicated by $Q_{\text{u}}$, $Q_{\text{b}}$, and $Q_{\text{f}}$, respectively (red dashed lines). C) Fraction of native contacts free-energy profiles $U(Q) = -k_{\rm{B}}T\log[P(Q)]$ for the two pH conditions. $P(Q)$ is the $Q$ probability density. Inset: conditional probabilities to be on a transition path $P(TP|Q)$ at position $Q$, suggesting that $Q$ is a good reaction coordinate for the all-atom $\alpha$3D trajectories under both pH conditions \cite{Hummer_2004}. For comparison, $Q$ is shifted by the peak value $Q_{\text{p}}$, often identified as the transition state, and rescaled by the distance between free energy minima $L_{\text{U}} = Q_{\text{f}} - Q_{\text{u}}$.
 \label{Fig_1}}
\end{figure}

$\alpha$3D is a designed, 73-residue, fast-folding protein \cite{Walsh_1999, Zhu_2003}, which folds into a tertiary structure comprised of three interacting $\alpha$-helical subunits (Fig.~\ref{Fig_1}A), and which exhibits distinct folded and unfolded states at neutral pH, both in experiment \cite{Chung_2011} and in simulation \cite{Lindorff_2011}. It has been shown that the folding kinetics of the $\alpha$3D protein are strongly influenced by non-native interactions \cite{Chung_2015, Best_2013}, in particular, by several non-native salt-bridge interactions between positively charged lysine and arginine side-groups and negatively charged aspartate and glutamate side groups. These salt bridges interact between the residue-chain segments associated with the distinct helices, even in the unfolded state \cite{Best_2013}. The folding kinetics of the $\alpha$3D protein were compared under low and neural pH conditions, using both experiments and all-atom simulations \cite{Chung_2015}. It was shown that lowering the pH significantly reduces both native and non-native salt-bridge interactions, therefore reducing the folded state stability but, surprisingly, increasing the folding rate. In the present article, we analyse the all-atom simulation data for the $\alpha$3D protein in both low and neural pH conditions (see Supplementary Materials Sec.~\RNum{1}) and show that the pH reduction perturbs the non-Markovian folding friction. Using the memory kernel extraction techniques we show that both the total friction and the non-Markovian memory time scale are significantly reduced when the pH is lowered, so much so that the non-Markovian features that dominate the $\alpha$3D folding kinetics at neutral pH are eliminated, allowing for a Markovian model description for the low pH folding kinetics. The same Markovian model fails to describe $\alpha$3D folding in neutral pH conditions, under-predicting the folding kinetics by a factor of 6, suggesting strong non-Markovian acceleration effects for the neutral pH system.

To track the folding progress of the $\alpha$3D protein, we project the atomic coordinates onto the fraction of native contacts reaction coordinate $Q$ (Fig.~\ref{Fig_1}B) \cite{Shakhnovich_1991, Best_2013} (Supplementary Materials Sec.~\RNum{2}). While the techniques employed in this article are, by definition, suitable for any arbitrary reaction coordinate, irrespective of the \textit{goodness} or \textit{poorness} of that reaction coordinate, we first note that $Q$ is equally good in both pH conditions, judged by the following criteria: for the neutral pH system, $Q$ exhibits a pronounced two-state distribution (Fig.~\ref{Fig_1}C) and a sharply peaked conditional transition-path probability $P(TP|Q) = P(Q|TP)P(TP)/P(Q)$ (Fig.~\ref{Fig_1}C inset and Supplementary Materials Sec.~\RNum{2}), i.e. the probability to be on a transition path when at $Q$ \cite{Hummer_2004}. $P(Q)$ and $P(TP)$ are the probability density for $Q$ and the probability of being on a transition path, respectively. $P(Q|TP)$ is the probability density conditional to being on a transition path. $P(TP|Q)$ provides a measure for the quality of a reaction coordinate, and it has been shown that $Q$ optimally reveals the transition state for proteins \cite{Best_2013}, which is apparent since $P(TP|Q)$ is sharply peaked and almost reaches a value of 0.5, which is an indication of a transition state in 1D Markovian theory. These results suggest that non-Markovian effects should be negligible \cite{Berezhkovskii_2018}, which we later show not to be the case.

Simulated folding times, as represented by the average residency time (the average time spent in a state before making a transition) in the unfolded states, of 10.2 $\pm \; 2 $ $\mu$s and 24.2 $\pm \; 9 $ $\mu$s in the low and neutral pH conditions, respectively, can be compared to 10 $\mu$s and 20 $\mu$s, as reported previously for the same data \cite{Chung_2015}, and experimental values of $\sim 6$ $\mu$s \cite{Zhu_2003} and $\sim 10$ $\mu$s \cite{Chung_2011}. Due to the strong asymmetry in the free energy profiles (Fig.~\ref{Fig_1}C) and the large differences between the folded-state energies, we here only consider transitions which connect the unfolded state minimum $Q_{\text{u}}$ to the barrier top $Q_{\text{b}}$. Therefore, we define the folding times $\tau^{\text{MD}}_{\text{MFP}}$ in the MD simulations as the mean first-passage times from $Q_{\text{u}}$ to $Q_{\text{b}}$. Measured in this way, we evaluate folding times of $\tau^{\text{MD}}_{\text{MFP}} = 5.2 \; \mu\rm{s}$ and $\tau^{\text{MD}}_{\text{MFP}} = 6.4 \; \mu\rm{s}$ for the low and neutral pH systems. These are significantly different to the residency times due to recrossing events, i.e., excursions that reach $Q_{\text{b}}$, but which do not continue to $Q_{\text{f}}$. In Supplementary Material Sec. \RNum{3}, we discuss the various definitions for folding times and recrossing, and we present folding time distributions and error estimates. By any definition of the folding time, we see that the low pH system folds slightly faster than the neutral system. As pointed out previously \cite{Chung_2015}, this is interesting since the folding barrier is slightly higher for the low pH system (Fig.~\ref{Fig_1}C), so the folding time reduction for low pH must be due to friction effects. 

\begin{figure}[t!]
\includegraphics[scale=1.1]{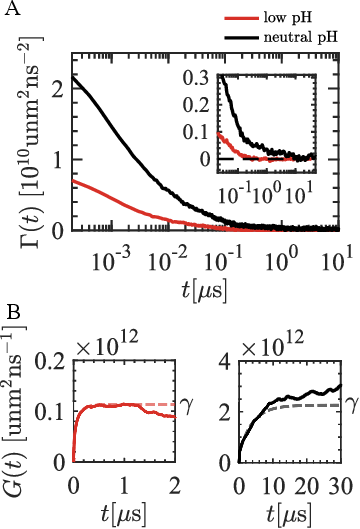}
\caption{Friction and memory effects in $\alpha$3D dynamics. A) Memory kernels $\Gamma(t)$, extracted for two different pH conditions. Inset: magnification of the long-time behaviour. B) Running integrals $G(t)$ for the two extracted memory kernels in A) (bold lines). The dashed lines show running integrals constructed from four-component exponential fits to the memory kernel data in A). Total friction $\gamma$ is given by the plateau value of $G(t\rightarrow\infty)$, taken from the fitted curves. \label{Fig_2}}
\end{figure}

To evaluate the friction, we map the $Q(t)$ trajectories for $\alpha$3D onto a generalised Langevin equation \cite{Zwanzig_1961, Mori_1965}, given by
\begin{equation}\label{GLE}
\begin{split}
m\ddot{Q}(t) &= -\int\limits_{0}^{t}\Gamma(t-t^{\prime})\dot{Q}(t^{\prime})dt^{\prime}  - \nabla U\big[Q(t)\big]+  \xi(t).
\end{split}
\end{equation}
$\Gamma(t)$ is the friction memory kernel, $U(Q)$ is the free energy profile (Fig.~\ref{Fig_1}C), and $m$ is the effective mass of $Q$, which is assumed to have no position dependence \cite{Dalton_2023}. $\xi(t)$ is the random force term satisfying the fluctuation-dissipation theorem $\langle \xi(t) \xi(t^{\prime}) \rangle =k_{\rm{B}}T\Gamma(t - t^{\prime})$, where $k_{\rm{B}}T$ is the thermal energy. Using recent memory kernel extraction techniques, we extract the running integral of the memory kernel $G(t)=\int_0^t \Gamma(t^{\prime})dt^{\prime}$ directly from the time series of $Q(t)$ \cite{Kowalik_2019, Ayaz_2021}, where the total friction $\gamma$ acting on $Q(t)$ is determined by the plateau value $\gamma=G(t\rightarrow\infty)$. In Fig.~\ref{Fig_2}A, we show the memory kernels for the $\alpha$3D protein under the two pH conditions (for additional information, see Supplementary Material Sec.~\RNum{4})). Lowering the pH in the MD simulations decreases both the amplitude and the decay time of the memory. The latter is evident in the long-time tails, as magnified in the inset. The memory kernel for neutral pH spans microseconds. However, for the low pH, the decay is over an order of magnitude faster. From the running integrals $G(t)$ in Fig.~\ref{Fig_2}B, we see that the total friction $\gamma$ is reduced by a factor of $\sim$23. Thus, whereas the amplitude of $\Gamma(t)$ is reduced by a factor of $\sim$2 at short times, the dramatic reduction in total friction is caused primarily by the reduced memory time.

\begin{figure}[b!]
\includegraphics[scale=0.72]{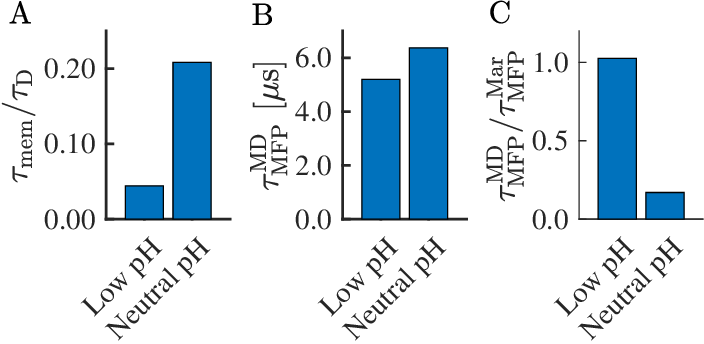}
\caption{Comparison of dynamic time scales for $\alpha$3D folding under two pH conditions. A) The ratio of the memory time $\tau_{\text{mem}}$ and diffusion time $\tau_{\text{D}}$ suggests that non-Markovian effects are more pronounced in the neutral pH condition. B) MD simulation mean first-passage times for transitions from unfolded state minimum to barrier top .C) Comparison of first mean-passage times $\tau^{\text{MD}}_{\text{MFP}}$ between $Q_{\text{u}}$ and $Q_{\text{b}}$ measured in simulations and a purely Markovian prediction of passage times $\tau^{\text{Mar}}_{\text{MFP}}$ (Eq.~\ref{MPT_Exact}). For the low pH condition, $\tau^{\text{MD}}_{\text{MFP}}$ is well predicted by a Markovian model. For the neutral pH system, the transition time is significantly faster than that measured in the MD simulation, suggesting non-Markovian acceleration effects. \label{Fig_3}}
\end{figure}

We evaluate the memory decay via the first moment of the memory kernel $\tau_{{\text{mem}}} = \int_{0}^{\infty} t \Gamma(t)dt/\int_{0}^{\infty} \Gamma(t)dt$, and obtain $\tau_{{\text{mem}}} = 3.4 \; \mu \text{s}$ and $0.07 \; \mu \text{s}$ for the neutral and low pH systems, respectively. Having extracted the friction, we can evaluate other characteristic time scales associated with the folding dynamics. The diffusion time $\tau_{\text{D}}=\gamma (Q_{\text{b}}-Q_{\text{u}})^2/k_{\rm{B}}T$ is the time to diffuse from the unfolded state to the barrier top in the absence of a free energy profile (a table of all relevant time scales and extracted system parameters is given in the Supplementary Materials Sec.~\RNum{1}). We expect memory effects to influence barrier crossing kinetics when $1{\times}10^{-2} < \tau_{\text{mem}}/\tau_{\text{D}} < 1{\times}10^1$ \cite{Kappler_2018}. This condition is well satisfied for the neutral pH system, but only just so for the low pH system (Fig.~\ref{Fig_3}A). Therefore, we expect non-Markovian effects to be more pronounced for the neutral pH system. 

In Fig.~\ref{Fig_3}B, we show the simulated folding mean-first passage times for the two pH systems. To quantify deviations from Markovianity, we predict transition times using the following Markovian overdamped formula which explicitly accounts for the free energy profiles given in Fig.~\ref{Fig_1}C \cite{Zwanzig_Book}: 
\begin{equation}\label{MPT_Exact} 
\begin{split}
\tau_{\rm{MFP}}^{\rm{\text{Mar}}} = \frac{\gamma}{k_{\rm{B}}T}\int\limits_{Q_\text{u}}^{Q_\text{b}} dx \text{e}^{ U(x)/k_{\rm{B}}T }
 \int\limits_{-\infty}^{x} dy\text{e}^{- U(y)/k_{\rm{B}}T}.
\end{split}
\end{equation}
Deviations between $\tau_{\rm{MFP}}^{\rm{Mar}}$ and $\tau_{\rm{MFP}}^{\rm{MD}}$ can be due to either neglecting position dependent friction, or the absence of memory effects. It was shown recently that a position dependent Markovian friction cannot consistently describe both folding and unfolding \cite{Ayaz_2021, Dalton_2023}, suggesting that the position-dependent Markovian friction model is not suitable for protein folding (note that this analysis was carried out for the neutral $\alpha$3D protein considered in this article \cite{Dalton_2023}). A comparison of $\tau_{\rm{MFP}}^{\rm{Mar}}$ and $\tau_{\rm{MFP}}^{\rm{MD}}$ for the two pH systems shows that Eq.~\ref{MPT_Exact} accurately predicts the transition time for the low pH system (Fig.~\ref{Fig_3}C). This is expected since the memory time scale approaches $1{\times}10^{-2}\tau_{\text{D}}$, so $\alpha$3D exhibits effectively Markovian folding, satisfying a model with position-independent, memoryless friction. Folding in the neutral pH system, however, is accelerated by a factor of 6 compared to the Markovian prediction (Fig.~\ref{Fig_3}C).

\begin{figure}[t!]
\includegraphics[scale=1.1]{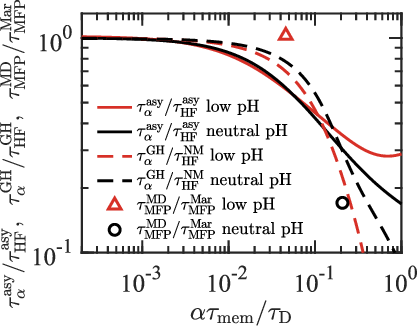}
\caption{Non-Markovian barrier crossing predictions over a range of memory time scales. Predictions by an asymptotic cross-over formula $\tau^{\rm{asy}}_{\alpha}/\tau^{\rm{asy}}_{\text{HF}}$ (solid lines) and the Grote-Hynes theory $\tau^{\rm{GH}}_{\alpha}/\tau^{\rm{GH}}_{\text{HF}}$ (dashed lines) are parametrised by a four-component exponential fit for the memory kernels in Fig.~\ref{Fig_2}A. All memory times are multiplied by a common factor $\alpha$ and each prediction is normalised by the respective high-friction Markovian limit ($\tau^{\rm{asy}}_{\text{HF}}$ and $\tau^{\rm{GH}}_{\text{HF}}$). The coloured symbols are for $\tau_{\rm{MD}}^{\rm{MFP}}/\tau_{\rm{Mar}}^{\rm{MFP}}$, located at $\alpha\tau_{\text{mem}}$ corresponding to $\alpha=1$. \label{Fig_4}}
\end{figure}

Barrier-crossing acceleration is a hallmark of non-Markovian effects and is predicted by various non-Markovian reaction rate theories. In Fig.~\ref{Fig_4}, we compare simulation results to various non-Markovian predictions, confirming that the neutral pH system folds rapidly, despite its significantly higher friction, due to memory speed-up effects. We parametrize both the Grote-Hynes (GH) theory \cite{Grote_1980} and a recent asymptotic cross-over formula \cite{Kappler_2018, Kappler_2019, Lavacchi_2020} by fitting memory kernels to $\Gamma(t) = \sum_{n=1}^4\gamma_n\text{e}^{-t/\tau_n}/\tau_n $ (Supplementary Materials Sec.~\RNum{4}) and uniformly scaling all memory times that enter each prediction by a common factor $\alpha$.

$\tau^{\rm{asy}}_{\alpha}/\tau^{\rm{asy}}_{\text{HF}}$ is the asymptotic cross-over formula \cite{Kappler_2019, Lavacchi_2020}, rescaled by the high-friction, or overdamped, Markovian limit $\tau^{\rm{asy}}_{\text{HF}}$ (Supplementary Material Sec.~\RNum{5}). This model was recently shown to be accurate for predicting the folding and unfolding times of a large population of fast-folding proteins \cite{Dalton_2023}, even capturing the non-Markovian barrier-crossing slow-down regime in some instances. Similarly, $\tau^{\rm{GH}}_{\alpha}/\tau^{\rm{GH}}_{\text{HF}}$ is the Grote-Hynes prediction, rescaled by the high-friction limit (Supplementary Material Sec.~\RNum{5}). In Fig.~\ref{Fig_4}, we show both predictions, plotted as a function of the 1st-moment memory time associated with the $\alpha$ scaling ($\alpha\tau_{\text{mem}}$). Both models predict reaction acceleration as $\tau_{\text{mem}}/\tau_{\text{D}}>1{\times}10^{-2}$. By including the values for $\tau_{\rm{MD}}^{\rm{MFP}}/\tau_{\rm{Mar}}^{\rm{MFP}}$ given in Fig.~\ref{Fig_3}C, we see that the deviations from Markovianity exhibited by the neutral pH protein system are expected since this system is deep in the memory-induced speed-up regime, as predicted by both non-Markovian models. Likewise, the low pH system aligns well with the Markovian limit, with deviations being likely due to the fact that both non-Markovian models incorporate harmonic approximations to the free energy profiles, which are not well represented for the two $\alpha$3D systems.

In summary, the extensive all-atom simulations of $\alpha$3D under two pH conditions \cite{Lindorff_2011, Chung_2015} have provided us with an ideal data set to explore the role of non-Markovian memory effects in protein folding dynamics. The reduction in pH, which physiochemically results in a reduction in the inter-residue salt-bridge interactions dispersed throughout the 73-residue chain, manifests as a direct perturbation on the friction acting on the coarse-grained reaction coordinate. By directly extracting the friction from the time series of the $Q$ reaction coordinate, we have shown that this perturbation modulates the amplitude and decay of the time-dependent friction kernel and that, in this instance, the time scale is reduced by so much that the pH=2 system behaves as an effectively Markovian system. For the neutral pH system, this is not the case and memory effects need to be explicitly accounted for to predict the dramatic decrease in folding time compared to that predicted by purely Markovian models.  \\

\noindent We acknowledge support by the ERC Advanced Grant No. 835117 NoMaMemo and the Deutsche Forschungsgemeinschaft (DFG) Grant No. SFB 1078. We thank the group of David E. Shaw for providing the all-atom simulation data analysed in this paper.

\bibliography{bib_file.bib}


\end{document}